\documentstyle[art10]{article} 
\newcommand{\epse}{\epsilon_e}
\newcommand{\epsp}{\epsilon_p}
\newcommand{\hrot}{h_{\rm rot}}
\newcommand{\hprec}{h_{\rm prec}}
\newcommand{\frot}{f_1}
\newcommand{\fprec}{f_2}
\newcommand{\omerot}{\omega_{\rm rot}}
\newcommand{\omeprec}{\omega_{\rm prec}}
\newcommand{\Hz}{{\rm Hz}}
\newcommand{\be}{\begin{equation}}
\newcommand{\ee}{\end{equation}}
\newcommand{\ba}{\begin{eqnarray}}
\newcommand{\ea}{\end{eqnarray}}
\newcommand{\heading}[1]{\noindent{\large\bf{#1}}} 
\renewenvironment{abstract}{\vskip 5mm\noindent {\bf Abstract.\,}\small}{} 
\renewcommand{\author}[1]{\vskip 3mm \noindent{\rm #1}\vskip 3mm} 
\newcommand{\address}[1]{\noindent{\small\it #1} \\} 

\def\lsim{~\rlap{$<$}{\lower 1.0ex\hbox{$\sim$}}} 
 
\def\gsim{~\rlap{$>$}{\lower 1.0ex\hbox{$\sim$}}} 
\def\@bibitem#1{\item\ifx#1\@empty\else\if@filesw \immediate\write\@auxout 
       {\string\bibcite{#1}{\the\value{\@listctr}}}\fi\fi\ignorespaces} 
\newenvironment{iapbib}[1]{\small\begin{thebibliography}{#1}\setlength{\itemsep}{-0.5mm}}{\end{thebibliography}} 
\newcommand{\mn}{MNRAS\,\,} 
\newcommand{\apj}{ApJ\,\,} 
\newcommand{\prd}{Phys.\ Rev.\ D\,\,}
\newcommand{\cqg}{Class.\ Quantum\ Grav.\,\,}
\begin{document} 
\heading{% 
%Begin Heading 
% 
Gravitational Waves from a Population of
Galactic Neutron Stars
% 
%End Heading 
}\footnote[2]{Paper presented at the 2nd Workshop on
Gravitational Wave Data Analysis, Orsay (France), 13-15
November 1997.} 
\par\medskip\noindent 
\author{Giacomo Giampieri}
\address{Queen Mary \& Westfield College,
Astronomy Unit, London E1 4NS} 
 
\begin{abstract} 
The existence of a large number of asymmetric, rotating
neutron stars, each individually emitting periodic or quasi-periodic 
gravitational waves in the frequency band around 100 Hz, raises the possibility
of detecting their combined signals, by exploiting the amplitude
modulation of the received waves as the antenna changes its orientation with
respect to fixed stars. This modulation is directly related to the amount of
anisotropy present in the source distribution, and, if detected, could give
valuable information about the spatial distribution of neutron stars 
in our Galaxy.
\end{abstract} 
\section{\large\bf Introduction} 
It is possible that a fraction of the
gravitational wave (GW) stochastic background is generated
astrophysically within our Galaxy, e.g., by a large population
of galactic sources emitting long-duration individual signals
which overlap in frequency. This is the case, for example,  in
the low-frequency region, where the expected signal
from unresolved close binary systems represents an effective
sensitivity limit for planned spaceborn interferometers, at
frequencies below $\sim 3$ mHz \cite{EVANS,BENDER}. 

A similar situation may arise for ground-based interferometers, 
when considering waves emitted by a large population of
non-axisymmetric, rotating neutron stars (NS).  This
contribution to the background signal, in any case, would not
have the same relevance as the galactic binaries background has
for space interferometers, for the following reasons: first, we
do not know how many rotating NS are present in our Galaxy, nor
how efficiently they emit GW, and therefore we do not know
whether their contribution to the background signal is relevant,
when compared to other stochastic sources. Also, one should
consider that the stochastic nature of this signal, in the sense
that individual sources cannot be resolved, results from the
finite duration of the observation time (i.e.\ from the finite
size of the resolution bin in the frequency spectrum), which in
the terrestrial case does not have any {\it a priori} upper
limit.  Finally,  even if one assumes that the number of sources
is large enough for the emitted waves to be unresolved in a given
integration time, one can always rely on a network of detectors,
which would allow disentangling this signal from other stochastic
sources. One can thus conclude that the overall signal from
galactic NS should not interfere with the search of other
sources, unless the NS birthrate is considerably higher than
today's estimates (see Eq.\ (\ref{eq:spectral}) below). 

It has been suggested, however, that detecting and measuring 
this signal with a single detector will provide direct and
valuable information about the spatial distribution of NS in the
Galaxy \cite{NS}, and in some cases could even be easier than
detecting individual sources \cite{GBG}. This detection
strategy is based on the fact that the antenna response is
not isotropic, and therefore the change of orientation of
the interferometer with respect to fixed stars generates an
amplitude modulation of the stochastic signal, given that the
sources are not distributed isotropically with respect to us. The
important point is that this modulation is totally
deterministic, and for a terrestrial interferometer it can be
easily expressed as the sum of four sinusoidal oscillations,
with frequencies $k/T_d\, (k=1,\dots,4)$, where $T_d$ is the
duration of the sidereal day \cite{antenna}. The amplitude of
these four periodic oscillations is related to the amount of
anisotropy in the  spatial distribution of sources, being
identically zero for a perfectly isotropic population (as
it should be, since in this case the detector's motion is
totally irrelevant). Thus, by measuring the modulations of
the variance of the stochastic signal one should be able to
obtain valuable information about how the NS are
distributed in our Galaxy \cite{NS}. 

This contribution summarizes some recent results on this subject. 
Full details of the calculations can be found in \cite{NS}.

\section{\large\bf The strength of the stochastic signal}

In general, deformations produced by internal forces in a
slowly-rotating NS are responsible for creating a slightly
non-axisymmetric configuration (described by the equatorial
oblateness $\epse$), and for making the star's angular velocity
precess around the near-symmetry axis (we call `wobble angle'
$\theta_W$ the constant angle between the rotation and symmetry
axes). Both effects result in the emission of sinusoidal
gravitational radiation, respectively at twice the rotation
frequency $\frot=\omerot/\pi$, and at its precessional sideband 
$\fprec=(\omerot+\omeprec)/(2\pi)\simeq \frot/2$. 
The  amplitudes of the two components are \cite{KT}
\ba
\hrot &=& \sqrt{32\over 5} \pi^2 \frac{\epse I \frot^2}{r}\,,\\
\hprec &=&  \sqrt{32\over 5} \pi^2\frac{\epsp \theta_W I
(2 \fprec)^2}{r} \,,
\ea
where $I$ is the star's moment of inertia, $\epsp$ is the poloidal
oblateness, mainly due to centrifugal forces, and $r$ is the
distance to the earth.

Each NS is slowly spinning down due to the loss of energy and
angular momentum in the form of electromagnetic and
gravitational waves. The exact evolution of the star's rotation
frequency, and consequently of the GW frequency and amplitude,
depends on which of the two emission mechanisms is dominant.
This also affects in a significant way the spectral properties
of the cumulative signal from the whole population of NS. For
example, for a given birthrate $\nu$ and a given observing time
$T_{obs}$, there exists a threshold frequency $f_c$ below which
the signal from the NS population behaves like a stochastic
signal, in the sense that one expects, on average, more than one
source per frequency resolution bin. It turns out that the
relatively fast evolution due to electromagnetic emission makes
this continuum frequency $f_c$ being located below the frequency
band of interest ($f_c < f_{\min} \simeq 20$ Hz), unless  
\be
\nu \gsim 2\times 10^3 \mbox{yr}^{-1},
\label{limit1}
\ee
a birthrate  considerably higher than the present estimate
$\nu_{est}\simeq 10^{-2} \mbox{yr}^{-1}$.

If, however, a significant fraction of sources are born with a small
magnetic field, so that GW radiation reaction dominates the (much
slower) evolution, then the limit (\ref{limit1}) drops to
\be
\nu \gsim 5.5\times 10^{-3} \mbox{yr}^{-1}\,.
\label{limit2}
\ee
In deriving eqs.\ (\ref{limit1}) and (\ref{limit2}) it has been
assumed a magnetic field strength $|\vec{B}|\simeq
10^{12}$ Gauss, an ellipticity $\epsilon\simeq 10^{-6}$, a
moment of inertia $I \simeq 10^{45}$ g cm$^2$, and $T_{obs}  =
1$ yr;  a more general expression for the minimum birthrate
$\nu$ can be found in \cite{NS}.

In view of the fact that unresolved NS are likely to be spinning down
by gravitational radiation reaction, we will consider only this
mechanism from now on. Also, for the sake of brevity, the precessional
component $\hprec$ of the radiation will be neglected, by imposing
$\theta_W=0$. The rotational component $\hrot$ gives, in the confusion
region $f\le f_c$, a stochastic signal of spectral amplitude 
\ba
S_h(f) &=& \frac{I \nu}{5 f} 
\int\limits_V \frac{\rho(r)}{r^2} d\vec r
\simeq 
1.5\times 10^{-51} \, \Hz^{-1}\, \left( f\over 100 \,\Hz\right)^{-1}
\nonumber\\
&\mbox{}&
\times \left(\nu\over 10^{-2}\, {\rm yr}^{-1}\right)
\left(I \over 10^{45} \, {\rm g~cm}^2 \right)\left[\int\limits_V 
\frac{\rho(r)}{r^2} d\vec r \over (10\, {\rm kpc})^{-2}\right]\,.
\label{eq:spectral}
\ea

Note that the spectral level (\ref{eq:spectral}) does not
depend on the oblateness of the sources, a direct consequence of
the fact that  gravitational radiation reaction is assumed to
be the dominant spindown mechanism. 

As one can infer from Eq.\ (\ref{eq:spectral}), a birthrate of
one pulsar each 100 years gives a signal well below the expected
sensitivity of planned terrestrial detectors. Unfortunately, the
actual rate is rather unknown, since a few hundred pulsars out
of perhaps $10^8$ rotating NS in our Galaxy have been detected,
and presumably the strongest sources have not yet been detected
electromagnetically. However, as shown in the next section,
this stochastic signal can be detected even if its strength is
much less than that of the instrumental noise, by exploiting the
peculiar time dependence of the signal's amplitude. 

\section{\large\bf Amplitude modulation of the stochastic
signal}

As already  mentioned in the Introduction, the stochastic signal
generated by any anisotropic distribution of sources, as seen by
an  interferometer anchored on the surface of the earth, is
non-stationary. It is now possible to quantify this effect, and
relate it to the actual distribution of sources via a simple
integral.

In particular, we will focus our attention to the signal's
expected variance, which can be expressed as
\be
\sigma^2(t) = \int \! df\, S(f,t)\,,
\ee
where $S(f,t)$ is the generalization of Eq.\ (\ref{eq:spectral})
which takes into account the actual response of the
interferometer, through its antenna pattern $P(\theta,\phi,t)$,
i.e. 
\be
S(f,t) = \frac{I \nu}{2 f} \int\limits_V \frac{P(\theta,\phi,t)
\rho(\vec{r})}{r^2} d\vec{r}\,.
\label{eq:spectral2}
\ee
Note that Eq.\ (\ref{eq:spectral2}) becomes identical to Eq.\
(\ref{eq:spectral}) if the spatial distribution is isotropic
with respect to us, since $\int\! d\Omega\, P(\theta,\phi,t)
\equiv 8 \pi/5$. For a terrestrial interferometer in
the equatorial reference frame one finds
\be
P(\theta,\phi,t) = \sum\limits_{k=0}^4 \left\{
a_k(\theta,\phi)\cos(k \omega_d t)
+ b_k(\theta,\phi) \sin(k \omega_d t)\right\}\,,
\label{eq:antenna}
\ee
where the Fourier coefficients $a_k, b_k$ are explicitly given
in \cite{antenna}, and $\omega_d$ is the earth sidereal frequency
$\omega_d \simeq 7\times 10^{-5}$ rad s$^{-1}$.
Inserting Eq.\ (\ref{eq:antenna}) in Eq.\ (\ref{eq:spectral2})
one trivially obtains, upon integration over frequency 
\be
\sigma^2(t) = \sum\limits_{k=0}^4 \left\{
\alpha_k \cos(k \omega_d t)
+ \beta_k \sin(k \omega_d t)\right\}\,,
\label{eq:variance}
\ee
where
$$
\alpha_k \propto \int\limits_{4\pi}\! d\Omega\, a_k(\theta,\phi)
\, \int\limits_0^\infty \! dr\, \rho(r,\theta,\phi)\,,
$$
and analogous for $\beta_k$, with $a_k$ replaced by
$b_k$. Thus, the variance of the stochastic signal is a
deterministic function of time; its Fourier decomposition
reveals four discrete frequencies, harmonics of the earth
sidereal frequency $\omega_d$. The amplitudes of these four
spectral lines depend on the particular form of $\rho$, i.e.\ on
the spatial distribution of the sources. By measuring them, one
can therefore obtain information about the global distribution of
sources in the Galaxy. 

The detection strategy could be as follows: the data record, of
duration $T_{obs}\gg T_d$, will be divided into several batches
of duration $\tau\ll T_d$. For each batch $J$ we obtain an
estimate $s_J^2$ of the variance $\sigma^2(t_J)$. According to
Eq.\ (\ref{eq:variance}), the variables $\{s_J^2\}$ are expected
to show four sinusoidal modulations, of known frequencies
($\omega_d$ and its first three harmonics) and phases (depending
on the initial orientation of the detector with respect to the
Galaxy). The amplitude of these oscillations can be estimated
with a simple Fourier analysis,  in the hypothesis that the
instrumental noise is stationary.  Note that, in order for this
search to be successful, it is not required that  $S_h >
S_{noise}$, since we are not directly comparing  the GW
stochastic signal against the instrumental noise; instead, we are
trying to detect a deterministic and periodic signal
$\sigma^2(t)$ buried in a stationary random noise of amplitude
$\sigma_{noise}^2$. The main problem in this measurement comes
from those  noise sources that, far from being stationary,
present the same diurnal variations as the gravitational signal,
due for example to temperature changes in the mirrors. This
issue has been addressed in great detail by Giazotto et al.\
\cite{GBG}, where a careful monitoring of the temperature
fluctuations in the  mirrors has been advocated.

In order to estimate the magnitude of the effect, and in
particular the amplitude of the periodic terms of
$\sigma^2(t)$ with respect to the constant one,
we have considered different galactic populations, like the
bulge, the disk, the halo, etc. Table 1 shows some results in
this direction: for each of the three interferometers presently
under construction -- the two LIGO detectors in the United States
\cite{LIGO}, and the VIRGO detector in Italy \cite{VIRGO} --, we
have estimated the  amplitudes 
$\gamma_k\equiv\sqrt{\alpha_k^2 + \beta_k^2}~(k=0,\dots,4)$
generated by, respectively, (i) a spherical halo, (ii) an
exponential disk, and (iii) a cluster of sources located in the
galactic center.

\begin{table}[bthp]
\begin{center}

%Table 1, Galactic population
%\begin{tabular}{||c||c|c|c||c|c|c||c|c|c||} \hline
\begin{tabular}{||c|ccc|ccc|ccc||} \hline
%	 &\multicolumn{9}{|c||}{}\\[-.35truecm]
%k & \multicolumn{9}{c||}{$\log(\gamma_k)$}               
	 &\multicolumn{9}{c||}{}\\[-.35truecm]
k & \multicolumn{9}{c||}{$\log(\gamma_k)$}  \\[.1truecm]\hline
%	 &\multicolumn{3}{|c||}{\bf Halo}
%  &\multicolumn{3}{|c||}{\bf Disk}
%	 &\multicolumn{3}{|c||}{\bf Bulge}
	 &\multicolumn{3}{c|}{\bf Halo}
  &\multicolumn{3}{c|}{\bf Disk}
	 &\multicolumn{3}{c||}{\bf Bulge}
%  \\[-.35truecm]
\\
\cline{2-10}
  &	L-1     & L-2	         &	V
  &	L-1     & L-2	         &	V
  &	L-1     & L-2	         &	V
  \\ 
\cline{2-10}
0	&	-3.96	         & -3.96             & -3.96  
  &	-0.53         	& -0.55             & -0.54
  &	-2.27	         & -2.25             & -2.28
\\
1 & -6.87          & -6.93             & -6.87 
  & -1.07          & -1.15             & -1.08
  & -2.32          & -2.34             & -2.29
\\
2 & -7.24          & -7.04             & -7.20
  & -1.42          & -1.25             & -1.41
  & -2.70          & -2.45             & -2.59
\\
3 & --          			& --             			& --    
  & -1.97          & -2.12             & -2.02 
  & -2.96          & -3.09             & -3.02 
\\
4 & --         	 		& --             			& -- 
  & -2.21          & -2.18             & -2.12
  & -3.22          & -3.20             & -3.13
\\[.1truecm]
\hline
\end{tabular}

\end{center}
\caption{\baselineskip=10pt
Summary of the  Fourier coefficients
$\gamma_k\equiv\protect\sqrt{\alpha_k^2+\beta_k^2}$, for various 
galactic components.
For each population, the columns marked
L-1, L-2, and V are obtained taking into account the
orientation  on the earth surface of, respectively, the two LIGO
detectors and the VIRGO detector. 
For further details, see
\protect\cite{NS}.}
\end{table}

From Table 1, one can conclude
that the largest contribution to the periodic terms of
$\sigma^2$ comes from the galactic disk. Also important is the
signal coming from a localized cluster of sources, like
the galactic bulge or even a rich globular cluster. The
periodic contributions from the spherical halo are expected to
be negligible.

%\acknowledgements{} 

% 
\vfill 
\end{document}